\documentclass{article}

\PassOptionsToPackage{numbers,sort&compress}{natbib}
\usepackage[preprint]{neurips_2026}

\usepackage[utf8]{inputenc} 
\usepackage[T1]{fontenc}    
\usepackage{hyperref}       
\usepackage{url}            
\usepackage{booktabs}       
\usepackage{amsfonts}       
\usepackage{nicefrac}       
\usepackage{microtype}      
\usepackage{xcolor}         
\usepackage{amsmath}
\usepackage{amssymb}
\usepackage{comment}
\usepackage{graphicx}
\usepackage{amsthm}
\theoremstyle{definition}
\newtheorem{definition}{Definition}
\usepackage{enumitem}

\usepackage{multirow}
\usepackage[table,dvipsnames]{xcolor}

\usepackage{caption}
\usepackage{algorithm}
\usepackage{algorithmic}

\title{From Bilinear to Linear: Differentially Private Federated LoRA via Low-Dimensional Parameterization}

\author{%
Lele Zheng \quad
Ruijie Hu \quad
Tao Zhang \quad
Ke Cheng \quad
Yulong Shen \\[5pt]
School of Computer Science and Technology, Xidian University, Xi'an, China \\[5pt]
}

\begin{document}

\maketitle

\begin{abstract}
Federated Low-Rank Adaptation (LoRA) provides an efficient solution for fine-tuning large language models across distributed and privacy-sensitive data.
However, despite avoiding raw data sharing, federated LoRA remains vulnerable to privacy leakage through transmitted model updates. 
Differential privacy (DP) mitigates such leakage, but integrating DP into federated LoRA introduces two fundamental challenges: aggregation mismatch from independently averaging low-rank factors, and quadratic noise amplification when noise is injected into both factors.
To address these challenges, we propose FedHSIP, a differentially private federated LoRA framework based on a unified low-dimensional parameterization. FedHSIP reformulates all LoRA parameters into a shared low-dimensional trainable vector, enabling clients to optimize and communicate only low-dimensional updates. This reformulation transforms federated LoRA from a bilinear factor aggregation problem into a unified linear parameter space, thereby eliminating aggregation mismatch and preventing the quadratic amplification of DP noise.
To further handle non-IID data, we introduce a heterogeneity- and sensitivity-aware isometric projection, constructed from warm-up statistics, which groups coordinates with compatible cross-client update patterns while balancing sensitivity, update energy, and heterogeneity across the low-dimensional space.
Extensive experiments on natural language understanding and generation benchmarks show that FedHSIP consistently outperforms existing federated LoRA methods under both private and non-private settings, achieving up to 3-4\% improvements under differential privacy while reducing communication cost by over 80\% and maintaining robustness under heterogeneous data distributions.
\end{abstract}

\section{Introduction}
Federated fine-tuning of pre-trained language models offers a promising paradigm for adapting foundation models to distributed \citep{yin2021comprehensive,li2020federated,zhang2026instruction}, privacy-sensitive data without centralizing raw client datasets. However, full-parameter federated fine-tuning is often prohibitive due to the substantial memory, computation, and communication costs imposed on local clients. Parameter-efficient fine-tuning methods, especially Low-Rank Adaptation (LoRA) \citep{hu2022lora}, have emerged as a practical alternative. By freezing the backbone model and training only a small set of low-rank adapters, LoRA substantially reduces the number of trainable parameters and the communication burden, making federated adaptation of large pre-trained models considerably more practical \citep{zhang2024towards,guo2024selective,sun2024improving,wang2024flora}.

Although federated learning prevents direct sharing of raw data \citep{mcmahan2017communication}, the transmitted gradients or model updates may still reveal sensitive information pertaining to local training samples. Differential privacy (DP) \citep{dwork2014algorithmic}, commonly instantiated via sample-level DP-SGD \citep{song2013stochastic,bassily2014private,abadi2016deep}, provides a rigorous framework to mitigate such leakage. Integrating DP into federated LoRA presents non-trivial challenges.
First, the direct averaging of local LoRA factors does not correspond to averaging the resulting model updates, resulting in aggregation mismatch and introducing undesirable cross-client interaction terms, particularly under non-IID data distributions. Second, when DP noise is injected into both LoRA factors, the noise components interact through the matrix product, producing a quadratic term that can substantially amplify the effective noise and degrade model utility \citep{kang2024federated,sun2024improving}. Existing approaches often address this issue by freezing one factor \citep{sun2024improving}; however, this strategy confines the update to a fixed subspace, thereby limiting the expressive capacity of the adapted model \citep{guo2024selective,zhang2023lora}. Alternative methods attempt to recover both factors via more elaborate server-side reconstruction, but this increases algorithmic complexity and remains constrained within the original bilinear parameterization \citep{lee2025fedsvd,wen2025differentially}.

To address the two core challenges identified in the preceding section, we adopt a low-dimensional parameterization strategy, transitioning federated LoRA from a bilinear factorization to a linear representation. Specifically, instead of treating LoRA as two coupled matrices that must be privately updated and aggregated separately, all LoRA parameters are flattened into a high-dimensional vector and reconstructed from a single trainable low-dimensional vector $\theta_d$ via a linear projection \citep{li2025uni}. This reformulation transforms both the optimization and communication process: clients optimize and transmit only $\theta_d$, while the server aggregates updates exclusively in the shared low-dimensional space. By operating on a single linear parameterization rather than the original matrix product, this approach eliminates aggregation mismatch and mitigates the quadratic amplification of DP noise, thereby fundamentally addressing the two principal obstacles of differentially private federated LoRA.

Low-dimensional parameterization requires an appropriate projection, which is critical for stable and high-quality federated updates under privacy and heterogeneous clients.
Random projections can combine high-dimensional coordinates that differ in direction, exhibit heterogeneous client-specific behavior, or have highly uneven sensitivity. This mixing can lead to destructive interference, substantially weakening the projected signal. When combined with gradient clipping and Gaussian perturbation imposed by DP, the useful signal can be further suppressed, undermining both optimization efficiency and model utility. Therefore, in the private federated context, a desirable projection should preserve global structure, maintain approximate isometry, and be aggregation-friendly while balancing sensitivity and heterogeneity across clients.

To this end, we propose a heterogeneity- and sensitivity-aware isometric projection (HSIP) for private federated LoRA. HSIP constructs the low-dimensional projection by leveraging statistical summaries collected during a preliminary warm-up phase, capturing the cross-client update patterns, sensitivity, and gradient magnitudes of the high-dimensional coordinates. Based on these summaries, coordinates are assigned to low-dimensional slots in a manner that minimizes destructive interference, balances sensitivity and update energy, and preserves approximate isometry. During the main training stage, the projection is fixed: each client reconstructs the full LoRA parameters from the shared $\theta_d$ vector, performs sample-level DP-SGD locally, and uploads only the privatized low-dimensional update. By aligning the low-dimensional geometry with the statistical characteristics of the clients’ updates, HSIP enables aggregation-friendly, noise-resilient optimization that maintains model utility under non-IID and differentially private conditions. Our contributions are summarized as follows:
\begin{itemize}[leftmargin=*]

    \item We formulate federated LoRA using a unified low-dimensional parameterization, converting the original bilinear factorization into a linear representation. This design addresses the two core challenges of private federated LoRA, namely aggregation mismatch and quadratic DP noise amplification.
    \item We introduce FedHSIP, a heterogeneity- and sensitivity-aware isometric projection that constructs the low-dimensional coordinate space from statistical summaries of client updates. This design ensures that each slot aggregates coordinates with compatible directions and balanced sensitivities, enabling aggregation-friendly and noise-resilient optimization. FedHSIP is fully compatible with sample-level differential privacy and remains robust to federated non-IID data.
    \item Extensive experiments show that FedHSIP consistently achieves strong performance across natural language understanding and generation benchmarks, remains robust under varying degrees of client data heterogeneity, and substantially reduces communication costs by exchanging only low-dimensional updates during the main federated training stage.
\end{itemize}

\section{Background}
\label{background}
\paragraph{Federated Learning with LoRA.}
The objective of Federated Learning with LoRA \citep{ye2024openfedllm,kuang2024federatedscope,zhang2024towards} is to minimize the following global objective:
\begin{equation}
F(W)=\frac{1}{K}\sum_{k=1}^{K} f_k(W_k,\mathcal{D}_k),
\end{equation}
where
\begin{equation}
f_k(W_k,\mathcal{D}_k)=\frac{1}{|\mathcal{D}_k|}\sum_{\xi\in \mathcal{D}_k} l(W_k,\xi)
\end{equation}
denotes the local objective function at client $k$. Here, $W_k$ represents the weight matrix of the local model, and $\mathcal{D}_k$ denotes the local dataset owned by client $k$. The function $l(W_k,\xi)$ computes the loss on an individual data sample $\xi\in \mathcal{D}_k$, while $f_k$ corresponds to the average loss over the entire local dataset. In LoRA, instead of directly training the full weight matrix $W_k$, the model is updated by adding a low-rank decomposition term to the pretrained weight $W_0$, and the adapter parameters are optimized via a gradient-based oracle. Specifically,
\begin{equation}
\frac{\partial l}{\partial A_k}=\frac{\alpha}{r}B_k^\top \frac{\partial l}{\partial W_k},
\qquad
\frac{\partial l}{\partial B_k}=\frac{\alpha}{r}\frac{\partial l}{\partial W_k}A_k^\top,
\end{equation}
where $A_k\in \mathbb{R}^{r\times n}$ and $B_k\in \mathbb{R}^{m\times r}$ are low-rank matrices of rank $r$ (typically $r\ll \min(m,n)$), and $\alpha$ is a scaling factor. The matrices $A_k$ and $B_k$ are precisely the parameters learned during fine-tuning.

A straightforward approach to federated LoRA is to allow each client to locally train its LoRA matrices ($A_k$ and $B_k$), followed by server-side average aggregation \citep{guo2024fedlfc,liu2025differentially}. \textbf{However, such direct averaging is inconsistent with the exact global model update \citep{wang2024flora}.} Specifically,
\begin{equation}
\bar{W}
=
\frac{1}{K}\sum_{k=1}^{K} W_k
=
\frac{1}{K}\sum_{k=1}^{K}\left(W_0+\frac{\alpha}{r}B_kA_k\right)
\neq
W_0+\frac{\alpha}{K^2 r}\sum_{k=1}^{K} B_k \sum_{k=1}^{K} A_k.
\end{equation}

This mismatch introduces undesirable cross-term noise, which becomes particularly pronounced in the presence of data heterogeneity across clients.

\paragraph{DP in LoRA.}
Language models often memorize portions of their training data, which may lead to the disclosure of private information contained in local clients' datasets \citep{carlini2021extracting,carlini2023quantifying}. Differential privacy \citep{dwork2014algorithmic} provides a formal privacy guarantee by limiting the influence of any single data point on the learned model, thereby mitigating such leakage risks.

\begin{definition}[$(\varepsilon,\delta)$-Differential Privacy]
A randomized algorithm \(M\) is said to satisfy \((\varepsilon,\delta)\)-differential privacy if, for any pair of neighboring datasets \(D\) and \(D'\) that differ in exactly one record, and for any measurable subset \(E\) of all possible outputs of \(M\), it holds that
$\Pr\!\big(M(D)\in E\big)\le e^{\varepsilon}\Pr\!\big(M(D')\in E\big)+\delta.$ Here, \(\varepsilon\) denotes the privacy budget, and \(\delta\) upper-bounds the probability that the privacy loss exceeds \(\varepsilon\), i.e., the probability that the DP guarantee may fail.
\end{definition}

Differentially private stochastic gradient descent (DP-SGD) \citep{abadi2016deep} is the most adopted approach for differentially private optimization. It enforces privacy by clipping per-sample gradients to a fixed $\ell_2$ norm bound and adding Gaussian noise to the aggregated gradients. However, directly applying DP-SGD to the standard LoRA update mechanism introduces nontrivial challenges. \textbf{When independent DP noise is injected into the gradients of the two factor matrices $A$ and $B$, the corresponding noise components interact in a quadratic manner during the construction of the LoRA update $\Delta W$.} Specifically,
\begin{equation}
W_0 + (B+\xi_B)(A+\xi_A)
= W_0 + BA + \xi_B A + B\xi_A + \xi_B\xi_A,
\end{equation}
where this cross-term interaction can lead to substantial noise amplification. Existing methods typically mitigate both aggregation bias and noise amplification by freezing one factor matrix while optimizing only the other. \citep{sun2024improving,kang2024federated}. Nevertheless, since such a strategy restricts the update to a fixed subspace, it may limit the expressive capacity of the fine-tuned model; moreover, employing an objective defined solely for $B$ is not well aligned with the task objective. Furthermore, FedSVD \citep{lee2025fedsvd} adopts a two-step update scheme: clients privately update the matrix $B$ with DP, while the server aggregates the updated $B$ to form the product $BA$, and then performs SVD reparameterization on $BA$ to realize updates for both $A$ and $B$. In contrast, FedASK \citep{wen2025differentially} employs a two-stage sketching-based aggregation procedure to update the global $A$ and $B$ simultaneously.

\section{Method}
\label{method}

\subsection{Preliminaries and Motivation}
\label{preliminaries and motivation}
\paragraph{Low-Dimensional LoRA Parameterization.}
Suppose the pretrained backbone contains $L$ LoRA-adapted modules. For the $\ell$-th module, let the LoRA parameters be denoted by
\begin{equation}
B^\ell \in \mathbb{R}^{m^\ell \times r_\ell}, \qquad
A^\ell \in \mathbb{R}^{r_\ell \times n^\ell},
\end{equation}
where $r_\ell$ is the LoRA rank. Following the unified view of Uni-LoRA\citep{li2025uni}, we row-flatten all LoRA parameters and concatenate them into a single high-dimensional vector
\begin{equation}
\theta_D
=
\operatorname{Concat}\!\Bigl(
\operatorname{vecrow}(B^1),\,
\operatorname{vecrow}(A^1),\,
\ldots,\,
\operatorname{vecrow}(B^L),\,
\operatorname{vecrow}(A^L)
\Bigr)
\in \mathbb{R}^D,
\end{equation}
where $D$ denotes the total number of LoRA parameters. Instead of directly optimizing $\theta_D$, we optimize a low-dimensional trainable vector $\theta_d \in \mathbb{R}^d$ with $d \ll D$, and reconstruct the full LoRA parameter vector through a linear projection
\begin{equation}
\theta_D = P\theta_d,
\qquad
P \in \mathbb{R}^{D \times d}.
\end{equation}

From this perspective, the design of the projection matrix $P$ and whether it remains frozen during training become key factors affecting adaptation performance.

\paragraph{How to design an ideal $P$?}
The above discussion indicates that an ideal projection matrix $P$ should preserve globality, load balancing, and isometry, as approximately achieved by random uniform projection. However, under federated non-IID training with sample-level DP-SGD, these structural properties are not sufficient. Since $P$ also determines how heterogeneous client updates are mixed, how clipping distorts useful signals, and how Gaussian noise affects coordinate-wise signal-to-noise ratios, a globally shared, balanced, and isometric projection can still be statistically suboptimal when heterogeneity and DP noise are taken into account.

For the projected per-sample gradient $g_d=P^\top g_D$, each slot $j$ satisfies
$
g_{d,j}=\frac{1}{\sqrt{n_j}}\sum_{i\in G_j}g_{D,i},
$
where $G_j$ is the set of assigned high-dimensional coordinates. Hence, a low-dimensional coordinate mixes multiple gradient components. When these components point in conflicting directions, useful signals cancel before optimization. This is particularly detrimental under DP-SGD, where subsequent clipping and Gaussian noise further obscure the already weakened signal.

Client heterogeneity further aggravates this issue. In non-IID federated learning, the same LoRA coordinate may have different importance, scale, or even sign across clients. Random grouping may therefore place coordinates with incompatible cross-client update patterns into the same slot, causing cancellation both within local training and during server aggregation. In the extreme case, a slot can have a small aggregated mean but large inter-client variance, making it particularly vulnerable to fixed-magnitude DP noise. Therefore, an effective projection should create low-dimensional coordinates that are not only compact, but also aggregation-friendly. 

Moreover, under DP-SGD, a slot’s statistical burden is determined not only by its size, but also by the sensitivity, gradient energy, and cross-client variability it aggregates. Grouping high-sensitivity or high-variance coordinates into the same slot can intensify clipping distortion and make the resulting update dominated by Gaussian noise.

\textbf{These observations motivate a projection that retains the globality, isometry, and load balancing, while additionally being heterogeneity-aware and sensitivity-aware.} The next section introduces the projection design.

\subsection{Proposed Projection: Heterogeneity- and Sensitivity-Aware Isometric Projection}

We now introduce the core projection module of FedHSIP, termed Heterogeneity- and Sensitivity-Aware Isometric Projection (HSIP). The key idea is to preserve the one-hot, column-normalized, and frozen projection skeleton, which maintains the desirable properties of globality, isometry, and load balancing, while adopting a statistics-aware assignment rule tailored to federated learning with sample-level DP-SGD.

\begin{figure*}[t]
  \centering
  \includegraphics[width=\textwidth]{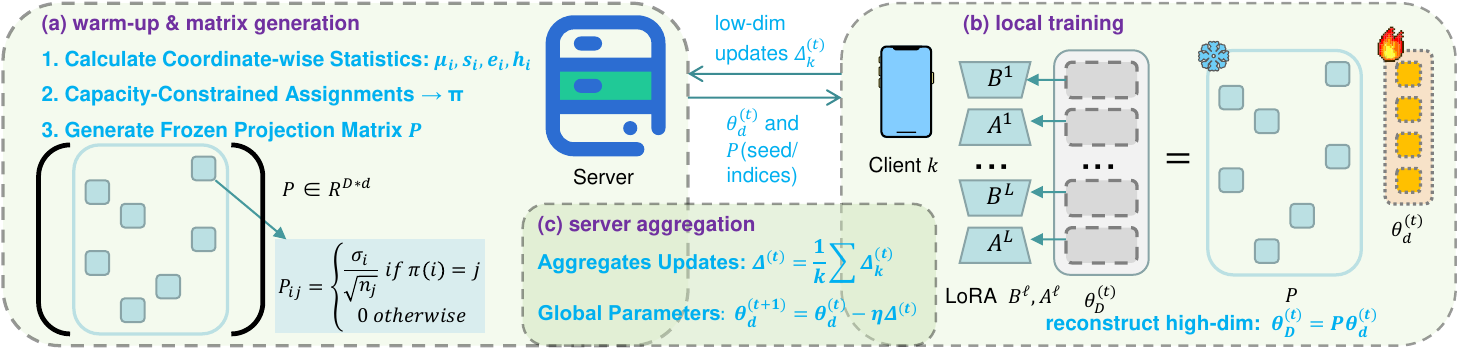}
  \caption{Overview of FedHSIP. (a) The server collects aggregated coordinate-wise statistics from clients and constructs a frozen sparse isometric projection $P$. (b) The server broadcasts $\theta_d^{(t)}$ and the projection rule; each client reconstructs the LoRA parameters by $\theta_D^{(t)} = P\theta_d^{(t)}$, performs sample-level DP-SGD locally, and uploads only the privatized low-dimensional update $\Delta_k^{(t)}$.}
  \label{fig:pipeline}
\end{figure*}

\subsubsection{Projection Parameterization}

Let $i = 1, \ldots, D$ index coordinates in the full LoRA parameter space and let $j = 1, \ldots, d$ index coordinates in the low-dimensional subspace. We define an assignment map
\begin{equation}
\pi : \{1, \ldots, D\} \to \{1, \ldots, d\},
\end{equation}
where $\pi(i) = j$ means that the $i$-th high-dimensional coordinate is assigned to the $j$-th low-dimensional slot. We further introduce a sign variable $\sigma_i \in \{-1, +1\}$, which is used for optional within-group sign alignment. Based on $\pi$ and $\sigma$, we define the projection matrix $P \in \mathbb{R}^{D \times d}$ as
\begin{equation}
P_{ij} =
\begin{cases}
\dfrac{\sigma_i}{\sqrt{n_j}}, & \pi(i) = j, \\
0, & \text{otherwise},
\end{cases}
\qquad
n_j = \left| \{ i : \pi(i) = j \} \right|.
\end{equation}
Here $n_j$ denotes the number of high-dimensional coordinates assigned to slot $j$. This construction preserves the one-hot and column-wise normalization structure. The mapping $\pi$ is constructed from federated warm-up statistics, and $\sigma_i$ is optionally chosen to reduce within-slot directional conflict. In practice, the dense matrix $P$ need not be explicitly materialized; it suffices to store the assignment indices, sign variables, and normalization factors, and reconstruct the projection implicitly during forward and backward passes.

\subsubsection{Warm-up Statistics Collection}

To make the projection aware of federated heterogeneity and DP-related sensitivity patterns, we introduce a warm-up stage before the main training phase. Rather than fixing the final projection at initialization, we first run several preliminary federated rounds under a provisional configuration and collect aggregated summaries that characterize each high-dimensional coordinate. The purpose of this stage is not to learn the full projection matrix directly, but to estimate which coordinates exhibit compatible update patterns and which coordinates should be prevented from concentrating in the same low-dimensional slot.

For each high-dimensional coordinate $i$, we construct a statistical profile consisting of four components. First, we define a client update profile $u_i \in \mathbb{R}^K$, which summarizes how the $i$-th coordinate behaves across clients. One possible choice is to let its $k$-th component be a standardized mean update of the form
\begin{equation}
u_{i,k}
=
\frac{\mathbb{E}_t\!\left[\Delta^{(t)}_{k,i}\right]}
{\sqrt{\mathbb{E}_t\!\left[\left(\Delta^{(t)}_{k,i}\right)^2\right] + \epsilon}},
\end{equation}
so that $u_i$ captures the relative direction pattern and cross-client importance of coordinate $i$. Second, we define a sensitivity statistic $s_i \geq 0$, which estimates how strongly this coordinate contributes to sample-level sensitivity, for example through per-sample gradient magnitude, pre-clipping second moments, or clipping frequency proxies. Third, we define an energy statistic $e_i \geq 0$, which measures update strength or importance, e.g., the average squared update magnitude. Fourth, we define a heterogeneity statistic $h_i \geq 0$, which measures cross-client variability, for example by the variance of the coordinate-wise mean update across clients.

We emphasize that the above forms are representative choices compatible with the design objective of HSIP. What matters is that they jointly describe, for each coordinate, its cross-client update pattern, its DP-related sensitivity burden, its signal strength, and its heterogeneity level. To avoid introducing additional privacy risks, these quantities are collected only through aggregated or privatized summaries, such as secure aggregation, noisy sketches, or low-precision coordinate statistics, rather than through raw coordinate-level disclosures from individual clients.

\subsubsection{Cost-Aware Balanced Assignment}

After the warm-up stage, we construct the projection by solving a balanced assignment problem over coordinates. Let
$
G_j = \{ i : \pi(i) = j \}
$
denote the set of high-dimensional coordinates assigned to slot $j$. The goal is to obtain an assignment that simultaneously reduces within-slot conflict, balances DP-related statistical load, and preserves the approximate count balance and isometric structure. Concretely, we define the objective
\begin{equation}
\min_{\pi,\sigma} \sum_{j=1}^{d}
\left[
\alpha\, \mathcal{C}(G_j,\sigma)
+ \beta\, (S_j - \bar{S})^2
+ \gamma\, (E_j - \bar{E})^2
+ \eta\, (H_j - \bar{H})^2
+ \lambda\, (|G_j| - D/d)^2
\right],
\end{equation}
where
$
S_j = \sum_{i \in G_j} s_i, E_j = \sum_{i \in G_j} e_i, H_j = \sum_{i \in G_j} h_i,
$
and $\bar{S}, \bar{E}, \bar{H}$ denote the corresponding average load per slot. The first term penalizes within-slot incompatibility, while the remaining terms encourage balanced sensitivity load, balanced signal energy, balanced heterogeneity load, and balanced occupancy, respectively.

For the conflict term, we use the client update profiles to measure how compatible two coordinates are when assigned to the same slot. Specifically, we define
\begin{equation}
\mathcal{C}(G_j,\sigma)
=
\sum_{\substack{i,i' \in G_j\\ i<i'}}
w_{ii'}
\left(
1 - \cos(\sigma_i u_i, \sigma_{i'} u_{i'})
\right),
\end{equation}
where $w_{ii'} \geq 0$ is an optional pairwise weight. This term is small when coordinates grouped into the same slot exhibit similar cross-client update patterns after sign alignment, and large when their update directions are inconsistent. Minimizing this term discourages destructive interference in the projected gradient and makes each slot more aggregation-friendly under federated optimization. Meanwhile, balancing $S_j$, $E_j$, and $H_j$ prevents a few slots from concentrating excessively sensitive, high-energy, or highly heterogeneous coordinates, which is important under clipping and Gaussian perturbation.

\subsection{Federated Optimization with HSIP}

Once the warm-up stage is completed, the assignment map $\pi$, sign vector $\sigma$, and the resulting HSIP projection matrix $P$ are fixed for the remainder of training, as shown in Fig~\ref{fig:pipeline}. The main training stage optimizes only the low-dimensional parameter vector $\theta_d$, while the full LoRA parameter vector is reconstructed through the frozen projection $\theta_D = P\theta_d$. In each communication round $t$, the server broadcasts the current global low-dimensional parameter $\theta_d^{(t)}$ together with the projection rule induced by $(\pi,\sigma)$. Each selected client then reconstructs the full LoRA parameter vector as
$
\theta_D^{(t)} = P\theta_d^{(t)},
$
maps it back to the corresponding layer-wise LoRA matrices $\{A^\ell, B^\ell\}_{\ell=1}^{L}$, and performs local training on its private dataset. After $E$ local update steps, client $k$ returns the low-dimensional model update
$
\Delta_k^{(t)} = \theta_{d,k}^{(t,\mathrm{final})} - \theta_d^{(t)}
$
to the server. The server aggregates the uploaded low-dimensional updates to obtain the next global iterate:
$
\theta_d^{(t+1)} = \sum_{k=1}^{K} p_k \bigl( \theta_d^{(t)} + \Delta_k^{(t)} \bigr),
$
where $p_k$ denotes the aggregation weight of client $k$, typically determined by the local sample size. Because all clients share the same frozen HSIP rule, aggregation is carried out in a common low-dimensional coordinate system, which preserves the communication efficiency and implementation simplicity while allowing the coordinate system itself to be adapted to heterogeneity and DP-related statistics through the warm-up stage.

\subsection{Complexity Analysis}
By maintaining a sparse one-hot and column-normalized projection backbone, HSIP enjoys favorable computational efficiency in the projection computation stage. Since each row of $P$ contains exactly one nonzero entry, $P$ has only $D$ nonzero entries in total. Therefore, reconstructing the full LoRA parameter vector $\theta_D=P\theta_d$ can be implemented in $O(D)$ time, rather than the $O(Dd)$ cost required by dense Gaussian projections. The storage overhead is also compact. Instead of materializing the dense matrix $P$, HSIP stores the learned low-dimensional vector $\theta_d$ together with the frozen projection rule specified by the assignment map $\pi$, sign vector $\sigma$, and normalization factors, which can be represented by index tensors and signs. The additional cost of HSIP comes from the warm-up stage and the construction of the projection from collected summaries. This overhead is incurred only once before federated training and does not affect the per-round communication object, which remains the low-dimensional parameter or update in $\mathbb{R}^d$.

\section{Experiments}
\label{experiments}

\subsection{Experimental Setups}
\label{experimental setups}
In this section, we evaluate the proposed method through a series of experiments. We first compare FedHSIP with state-of-the-art federated fine-tuning methods on the GLUE benchmark, including FedAvg, FFA-LoRA, FedSVD, and FedASK. We then extend our analysis to natural language generation tasks based on the GPT-2 model.

\subsection{Main Results}

\paragraph{Natural language understanding.}
We first evaluate FedHSIP on GLUE natural language understanding tasks under both non-private training and DP-SGD with $\epsilon \in \{3,6\}$ and $\delta=10^{-5}$. As shown in Table~\ref{tab:glue}, FedHSIP achieves the best non-private average accuracy of 87.76, improving over FedSVD and FFA-LoRA by 1.10 and 2.03 percentage points, respectively. This indicates that the proposed projection preserves standard utility despite optimizing and communicating only the low-dimensional vector $\theta_d$. The advantage becomes more evident under differential privacy. With $\epsilon=6$, FedHSIP obtains the best result across all five evaluation columns and reaches an average accuracy of 83.25, outperforming FedSVD by 3.40 points. Under the stricter budget $\epsilon=3$, FedHSIP remains the best-performing method, achieving an average accuracy of 82.87 and exceeding FedSVD by 4.44 points. The small degradation from $\epsilon=6$ to $\epsilon=3$ further suggests that HSIP provides an aggregation-friendly and noise-resilient low-dimensional training space for private federated LoRA.
\begin{table}[t]
    \caption{Results on five GLUE tasks under non-private and differentially private settings. For DP experiments, we use DP-SGD with $\epsilon \in \{3,6\}$ and $\delta=10^{-5}$. The best and second-best results are highlighted in \textbf{bold} and \underline{underline}, respectively.}
    \label{tab:glue}
    \centering
    \small
    \setlength{\tabcolsep}{4pt}
    \renewcommand{\arraystretch}{0.95}
    \begin{tabular}{c|l|cccccc}
        \toprule
        {\textbf{DP}} & \multirow{2}{*}{\textbf{Method}} & \multicolumn{2}{c}{\textbf{MNLI}} & \multirow{2}{*}{\textbf{SST-2}} & \multirow{2}{*}{\textbf{QQP}} & \multirow{2}{*}{\textbf{QNLI}} & \multirow{2}{*}{\textbf{Average}} \\
        \textbf{Budget} & & \textbf{Matched} & \textbf{Mismatched} & & & & \\
        \midrule

        \multirow{5}{*}{Non Private}
        & FedAvg
        & \underline{85.01}
        & \underline{85.50}
        & 50.92
        & \underline{86.89}
        & 50.56
        & 71.78
        \\

        & FFA-LoRA
        & 83.02
        & 83.42
        & \underline{93.00}
        & 84.54
        & 84.66
        & 85.73
        \\

        & FedSVD
        & 83.50
        & 83.79
        & 92.55
        & 85.32
        & \underline{88.14}
        & \underline{86.67}
        \\

        & FedASK
        & \textbf{85.42}
        & \textbf{85.65}
        & 50.92
        & \textbf{87.13}
        & 50.74
        & 71.97
        \\

        & \cellcolor{cyan!10}{FedHSIP}
        & \cellcolor{cyan!10}83.94
        & \cellcolor{cyan!10}84.08
        & \cellcolor{cyan!10}\textbf{94.04}
        & \cellcolor{cyan!10}85.65
        & \cellcolor{cyan!10}\textbf{91.09}
        & \cellcolor{cyan!10}\textbf{87.76}
        \\

        \midrule

        \multirow{5}{*}{\scalebox{1.3}{$\epsilon=6$}}
        & FedAvg
        & 75.16
        & 75.64
        & 82.57
        & 79.21
        & 53.49
        & 73.21
        \\

        & FFA-LoRA
        & 66.61
        & 68.01
        & \underline{91.17}
        & \underline{80.39}
        & 80.43
        & 77.32
        \\

        & FedSVD
        & 74.95
        & 76.00
        & 88.42
        & 79.38
        & \underline{80.51}
        & \underline{79.85}
        \\

        & FedASK
        & \underline{75.96}
        & \underline{76.45}
        & 80.16
        & 79.25
        & 61.41
        & 74.65
        \\

        & \cellcolor{cyan!10}{FedHSIP}
        & \cellcolor{cyan!10}\textbf{78.93}
        & \cellcolor{cyan!10}\textbf{79.92}
        & \cellcolor{cyan!10}\textbf{92.20}
        & \cellcolor{cyan!10}\textbf{80.42}
        & \cellcolor{cyan!10}\textbf{84.77}
        & \cellcolor{cyan!10}\textbf{83.25}
        \\

        \midrule

        \multirow{5}{*}{\scalebox{1.3}{$\epsilon=3$}}
        & FedAvg
        & \underline{73.61}
        & 74.37
        & 79.01
        & 77.84
        & 57.39
        & 72.44
        \\

        & FFA-LoRA
        & 53.96
        & 56.38
        & \underline{91.06}
        & \underline{79.85}
        & \underline{78.11}
        & 71.87
        \\

        & FedSVD
        & 72.79
        & 74.51
        & 88.76
        & 78.29
        & 77.80
        & \underline{78.43}
        \\

        & FedASK
        & 73.09
        & \underline{74.89}
        & 82.91
        & 78.77
        & 62.84
        & 74.50
        \\

        & \cellcolor{cyan!10}{FedHSIP}
        & \cellcolor{cyan!10}\textbf{78.37}
        & \cellcolor{cyan!10}\textbf{79.54}
        & \cellcolor{cyan!10}\textbf{92.09}
        & \cellcolor{cyan!10}\textbf{79.94}
        & \cellcolor{cyan!10}\textbf{84.42}
        & \cellcolor{cyan!10}\textbf{82.87}
        \\

        \bottomrule
    \end{tabular}
    
\vspace{-0.05in}
\end{table}

\paragraph{Heterogeneity of the data distribution ($\alpha$).}
We further evaluate the robustness of FedHSIP under non-i.i.d. data by partitioning MNLI across clients with Dirichlet concentration parameters $\alpha \in \{0.1,0.2,0.3,0.4,0.5\}$, where smaller $\alpha$ indicates stronger heterogeneity. Under the same RoBERTa-base DP-SGD setting with $\epsilon=6$ and $\delta=10^{-5}$, Table~\ref{tab:dirichlet} shows that FedHSIP achieves the highest average accuracy across all heterogeneity levels, reaching 75.62 and improving over FedASK and FedSVD by +2.96 and +3.64 percentage points, respectively. For $\alpha \in \{0.2,0.3,0.4,0.5\}$, FedHSIP consistently outperforms all baselines, with a particularly large gain at $\alpha=0.2$. When $\alpha=0.1$, the data distribution becomes extremely skewed and FedHSIP is not the best method, but it still surpasses FedAvg and FFA-LoRA. These results indicate that HSIP is especially effective when heterogeneous clients still share useful update structure, as its statistics-aware projection constructs a more aggregation-friendly and noise-resilient low-dimensional space.
\begin{table}
\caption{Results on MNLI with RoBERTa-base under varying client data heterogeneity. The data are partitioned by Dirichlet distributions with concentration parameter $\alpha \in \{0.1,0.2,0.3,0.4,0.5\}$, where smaller $\alpha$ denotes stronger non-IID heterogeneity.}
\label{tab:dirichlet}
\centering
\small
\begin{tabular}{l|ccccc}
\hline
\toprule
\multirow{2}*{\bf Method} & \multicolumn{5}{c}{\textbf{MNLI}} \\
       & $\alpha=0.1$ & $\alpha=0.2$ & $\alpha=0.3$ & $\alpha=0.4$ & $\alpha=0.5$\\
\midrule
$\text{FedAvg}$  & 63.92 & 55.79 & 60.19 & 72.78 & 75.16  \\
$\text{FFA-LoRA}$ & 67.05 & 67.09 & 59.32 & 62.32 & 66.61  \\
$\text{FedSVD}$ & \textbf{70.71} & 70.25 & 71.22 & 72.79 & 74.94  \\
$\text{FedASK}$ & \underline{70.18} & \underline{70.52} & \underline{72.39} & \underline{74.25} & \underline{75.95}  \\
\rowcolor{cyan!10}$\text{FedHSIP}$ & 68.24 & \textbf{77.58} & \textbf{76.17} & \textbf{77.18} & \textbf{78.93}  \\
\bottomrule
\end{tabular}
\end{table}

\paragraph{Natural language generation.}
We further evaluate FedHSIP on GPT-2 using the E2E NLG Challenge under the non-private setting. As shown in Table~\ref{tab:gpt2_results}, FedHSIP achieves the best performance on four out of five metrics at rank $r=4$, obtaining 41.80 BLEU, 6.33 NIST, 38.37 MET, and 1.93 CIDEr. It also attains a competitive ROUGE-L score of 56.69, only 0.08 lower than the best baseline. Compared with the strongest competing method on each metric, FedHSIP improves BLEU, NIST, MET, and CIDEr by +0.13, +0.02, +0.14, and +0.02, respectively. These results show that the proposed low-dimensional parameterization and HSIP projection generalize beyond discriminative NLU tasks and preserve strong generation utility while maintaining the communication-efficient $\theta_d$-only federated training pipeline.

\begin{table}[t]
\centering

\begin{minipage}[t]{0.62\textwidth}
    \captionsetup{type=table}
    \caption{Results with GPT-2 on the E2E NLG Challenge, comparing various federated LoRA methods at rank $r=4$. 
    There are $3$ local epochs before every aggregation round. 
    }
    \label{tab:gpt2_results}
    \centering
    \scriptsize
    \setlength{\tabcolsep}{3pt}
    \resizebox{\linewidth}{!}{
    \begin{tabular}{lccccc}
    \toprule
    \multirow{2}{*}{\bf Method} & \multicolumn{5}{c}{\textbf{E2E NLG Challenge}} \\
     & BLEU $\uparrow$ & NIST $\uparrow$ & MET $\uparrow$ & ROUGE-L $\uparrow$ & CIDEr $\uparrow$ \\
    \midrule
    $\text{FedAvg}$   & \underline{41.67} & \underline{6.31} & \underline{38.23} & \textbf{56.77} & \underline{1.91} \\
    $\text{FFA-LoRA}$ & 40.83 & 6.25 & 38.03 & 55.75 & 1.86 \\
    $\text{FedSVD}$   & 41.14 & 6.27 & 37.95 & 56.01 & 1.87 \\
    $\text{FedASK}$   & 41.44 & \underline{6.31} & 38.21 & 56.43 & 1.90 \\
    \rowcolor{cyan!10}$\text{FedHSIP}$ & \textbf{41.80} & \textbf{6.33} & \textbf{38.37} & \underline{56.69} & \textbf{1.93} \\
    \bottomrule
    \end{tabular}}
\end{minipage}
\hfill
\begin{minipage}[t]{0.35\textwidth}
    \captionsetup{type=table}
    \caption{ \textbf{Communication cost per round} when using RoBERTa-base and LoRA with rank $r=8$.}
    \label{tab:communication}
    \centering
    \small
    \setlength{\tabcolsep}{3pt}
    \renewcommand{\arraystretch}{1.1}
    \begin{tabular}{lcc}
        \toprule
        \textbf{Method} & \textbf{Upload} & \textbf{Broadcast} \\
        \midrule
        FedAvg & 294,912 & 294,912  \\
        FFA-LoRA  & \underline{147,456}  & \underline{147,456} \\
        FedSVD & \underline{147,456}  & 294,912 \\
        FedASK & 442,368  & 516,096 \\
        \midrule
        \rowcolor{cyan!10}\text{FedHSIP} & \textbf{46,080} & \textbf{46,080} \\
        \bottomrule
    \end{tabular}
\end{minipage}

\end{table}

\paragraph{Communication cost.} 
Table~\ref{tab:communication} reports the per-round communication cost on RoBERTa-base with LoRA rank $r=8$. Since FedHSIP performs the main training stage entirely in the shared low-dimensional space, each client uploads only a privatized update in $\mathbb{R}^d$, and the server broadcasts only the updated low-dimensional parameter. As a result, FedHSIP requires only 46,080 parameters for both upload and broadcast, which is substantially lower than FedAvg, FFA-LoRA, FedSVD, and FedASK. In total per-round communication, FedHSIP exchanges 92,160 parameters, reducing the cost by 84.4\% compared with FedAvg, 68.8\% compared with FFA-LoRA, 79.2\% compared with FedSVD, and 90.4\% compared with FedASK. This reduction is achieved without freezing any LoRA factor or transmitting additional high-rank correction terms, because the HSIP projection rule is constructed once during warm-up and remains fixed throughout the main training stage. These results show that FedHSIP provides a more communication-efficient pipeline for private federated LoRA.

\subsection{Ablation Study}
\begin{figure}[t]
    \centering

    \begin{minipage}[t]{0.68\textwidth}
        \vspace{0pt}
        \centering
        \includegraphics[width=\linewidth,height=5.6cm]{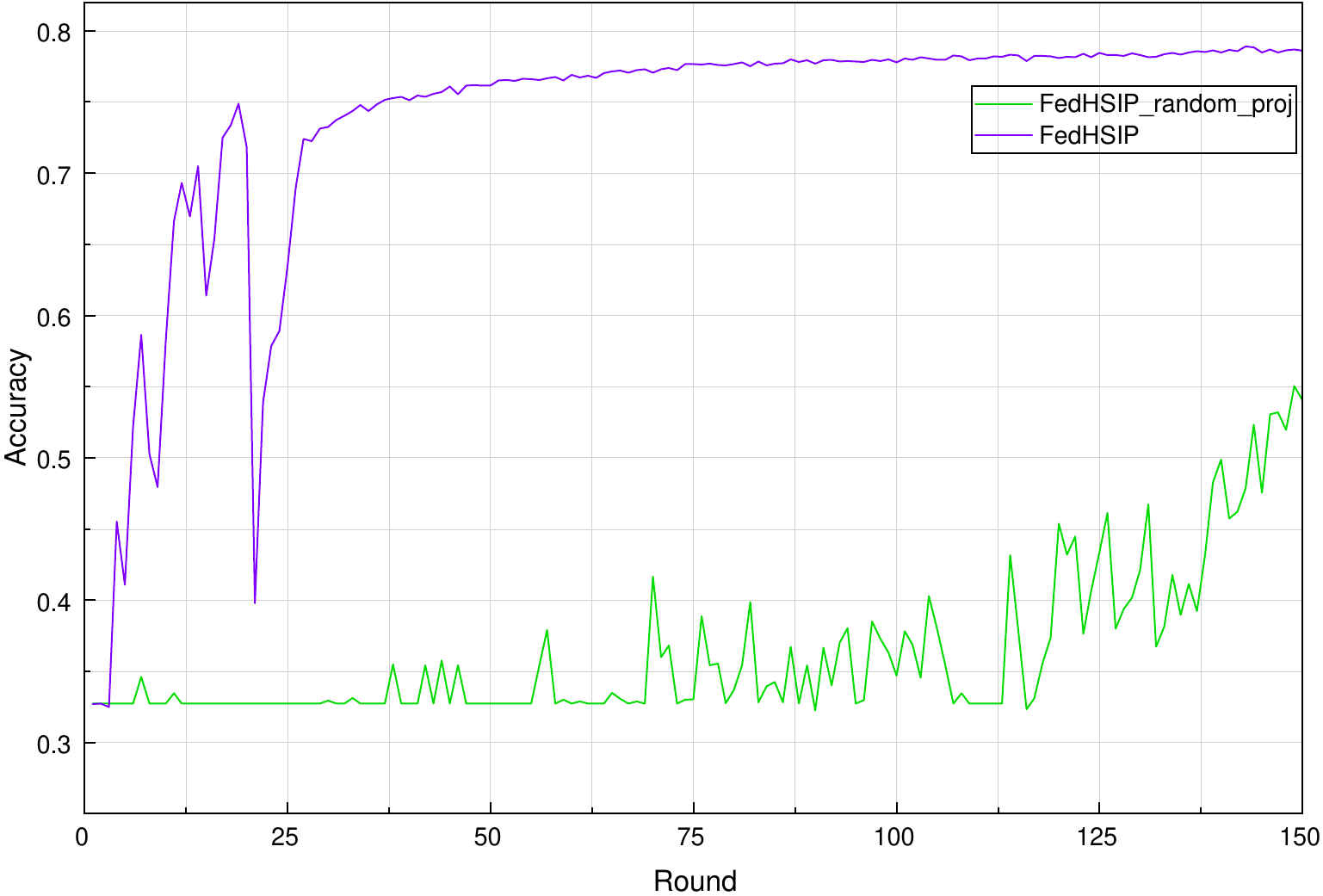}
        \captionof{figure}{Ablation study comparing HSIP with a random uniform projection using RoBERTa-base on MNLI with $\epsilon=6$ and $\delta=10^{-5}$.}
        \label{fig:ablation}
    \end{minipage}
    \hfill
    \begin{minipage}[t]{0.28\textwidth}
        \vspace{0pt}
        \centering
        \captionof{table}{\small Ablation study on the assignment coefficients of HSIP using RoBERTa-base on MNLI under DP-SGD with $\epsilon=6$ and $\delta=10^{-5}$. FedHSIP sets all coefficients $\alpha,\beta,\eta,\gamma,\lambda$ to $1$, while each variant increases one coefficient to $10$ and keeps the others unchanged.}
        \label{tab:ablation2}

        \vspace{2mm}
        \footnotesize
        \setlength{\tabcolsep}{4pt}
        \begin{tabular}{lc}
            \toprule
            \textbf{Method} & \textbf{Acc.}\\
            \midrule
            FedHSIP-alpha-10 & 79.63 \\
            FedHSIP-beta-10  & 79.04 \\
            FedHSIP-eta-10 & 79.39 \\
            FedHSIP-gamma-10 & 79.39 \\
            FedHSIP-lambda-10 & 79.32 \\
            \midrule
            \rowcolor{cyan!10}FedHSIP & 78.93 \\
            \bottomrule
        \end{tabular}
    \end{minipage}
\end{figure}

\paragraph{Comparison with random uniform projections.} 
To isolate the effect of the proposed heterogeneity- and sensitivity-aware assignment, we replace HSIP with a random uniform projection while keeping the rest of the training pipeline unchanged. Experiments are conducted on MNLI with RoBERTa-base under DP-SGD with $\epsilon=6$ and $\delta=10^{-5}$. As shown in Figure~\ref{fig:ablation}, the random projection exhibits unstable optimization and reaches 0.5505 accuracy, whereas FedHSIP reaches 0.7893, yielding a gain of 23.88 percentage points. This large gap shows that a global, count-balanced, and approximately isometric random projection is insufficient for private federated LoRA. Under non-IID data and DP-SGD, random grouping may mix coordinates with incompatible client update patterns or uneven sensitivity, weakening useful signals before clipping and Gaussian noise are applied. By contrast, HSIP uses warm-up statistics to group compatible coordinates and balance sensitivity, energy, and heterogeneity across slots, thereby providing a more aggregation-friendly and noise-resilient low-dimensional optimization space.

\paragraph{Ablation on assignment coefficients.} 
We examine the sensitivity of HSIP to the coefficients in the balanced assignment objective. Starting from the default FedHSIP setting where all coefficients are set to 1, we construct five variants by increasing one coefficient among $\alpha$, $\beta$, $\eta$, $\gamma$, and $\lambda$ to 10 while keeping the others unchanged. As shown in Table~\ref{tab:ablation2}, FedHSIP remains stable under all coefficient perturbations. All variants slightly improve over the default accuracy of 78.93, with results ranging from 79.04 to 79.63. The best performance is achieved by increasing $\alpha$, which controls the within-slot conflict penalty, suggesting that reducing directional incompatibility among grouped coordinates is particularly important under DP-SGD. Increasing $\eta$, $\gamma$, and $\lambda$ also yields consistent gains, confirming the usefulness of balancing heterogeneity, signal energy, and slot occupancy. The smaller but positive gain from increasing $\beta$ indicates that sensitivity balancing is helpful but less dominant in this setting. Overall, these results show that the default coefficient choice is already strong and that each term in the HSIP objective contributes to constructing a robust low-dimensional projection.

\section{Conclusion}
In this paper, we introduced FedHSIP, a differentially private federated LoRA framework that replaces bilinear factor aggregation with a shared low-dimensional parameterization. By optimizing and communicating only a single trainable low-dimensional vector, FedHSIP avoids factor-wise aggregation mismatch and mitigates the quadratic interaction of DP noise inherent in LoRA factor products. It further leverages a heterogeneity- and sensitivity-aware isometric projection, constructed from warm-up statistics, to enable robust optimization under non-IID and private federated training conditions. Extensive experiments demonstrate that FedHSIP improves utility under differential privacy, maintains robustness to client heterogeneity, and substantially reduces communication overhead. These results highlight the role of projection geometry in private federated fine-tuning. 

\clearpage
\bibliographystyle{plainnat}
\bibliography{neurips_2026}

@inproceedings{carlini2021extracting,
  title={Extracting training data from large language models},
  author={Carlini, Nicholas and Tramer, Florian and Wallace, Eric and Jagielski, Matthew and Herbert-Voss, Ariel and Lee, Katherine and Roberts, Adam and Brown, Tom and Song, Dawn and Erlingsson, Ulfar and others},
  booktitle={30th USENIX security symposium (USENIX Security 21)},
  pages={2633--2650},
  year={2021}
}

@article{carlini2023quantifying,
  title={Quantifying memorization across neural language models, 2023},
  author={Carlini, Nicholas and Ippolito, Daphne and Jagielski, Matthew and Lee, Katherine and Tramer, Florian and Zhang, Chiyuan},
  journal={URL https://arxiv. org/abs/2202.07646},
  volume={2202},
  year={2023}
}

@article{dwork2014algorithmic,
  title={The algorithmic foundations of differential privacy},
  author={Dwork, Cynthia and Roth, Aaron and others},
  journal={Foundations and trends{\textregistered} in theoretical computer science},
  volume={9},
  number={3--4},
  pages={211--407},
  year={2014},
  publisher={Now Publishers, Inc.}
}

@article{lee2025fedsvd,
  title={FedSVD: Adaptive Orthogonalization for Private Federated Learning with LoRA},
  author={Lee, Seanie and Park, Sangwoo and Lee, Dong Bok and Wagner, Dominik and Seong, Haebin and Bocklet, Tobias and Lee, Juho and Hwang, Sung Ju},
  journal={arXiv preprint arXiv:2505.12805},
  year={2025}
}

@article{sun2024improving,
  title={Improving lora in privacy-preserving federated learning},
  author={Sun, Youbang and Li, Zitao and Li, Yaliang and Ding, Bolin},
  journal={arXiv preprint arXiv:2403.12313},
  year={2024}
}

@article{wen2025differentially,
  title={Differentially Private Federated Low Rank Adaptation Beyond Fixed-Matrix},
  author={Wen, Ming and Zhu, Jiaqi and Xu, Yuedong and Zhou, Yipeng and Han, Dingding},
  journal={arXiv preprint arXiv:2507.09990},
  year={2025}
}

@inproceedings{ye2024openfedllm,
  title={Openfedllm: Training large language models on decentralized private data via federated learning},
  author={Ye, Rui and Wang, Wenhao and Chai, Jingyi and Li, Dihan and Li, Zexi and Xu, Yinda and Du, Yaxin and Wang, Yanfeng and Chen, Siheng},
  booktitle={Proceedings of the 30th ACM SIGKDD conference on knowledge discovery and data mining},
  pages={6137--6147},
  year={2024}
}

@inproceedings{kuang2024federatedscope,
  title={Federatedscope-llm: A comprehensive package for fine-tuning large language models in federated learning},
  author={Kuang, Weirui and Qian, Bingchen and Li, Zitao and Chen, Daoyuan and Gao, Dawei and Pan, Xuchen and Xie, Yuexiang and Li, Yaliang and Ding, Bolin and Zhou, Jingren},
  booktitle={Proceedings of the 30th ACM SIGKDD Conference on Knowledge Discovery and Data Mining},
  pages={5260--5271},
  year={2024}
}

@inproceedings{zhang2024towards,
  title={Towards building the federatedgpt: Federated instruction tuning},
  author={Zhang, Jianyi and Vahidian, Saeed and Kuo, Martin and Li, Chunyuan and Zhang, Ruiyi and Yu, Tong and Wang, Guoyin and Chen, Yiran},
  booktitle={ICASSP 2024-2024 IEEE international conference on acoustics, speech and signal processing (ICASSP)},
  pages={6915--6919},
  year={2024},
  organization={IEEE}
}

@article{wang2024flora,
  title={Flora: Federated fine-tuning large language models with heterogeneous low-rank adaptations},
  author={Wang, Ziyao and Shen, Zheyu and He, Yexiao and Sun, Guoheng and Wang, Hongyi and Lyu, Lingjuan and Li, Ang},
  journal={Advances in Neural Information Processing Systems},
  volume={37},
  pages={22513--22533},
  year={2024}
}

@inproceedings{guo2024fedlfc,
  title={Fedlfc: Towards efficient federated multilingual modeling with lora-based language family clustering},
  author={Guo, Zhihan and Zhang, Yifei and Zhang, Zhuo and Xu, Zenglin and King, Irwin},
  booktitle={Findings of the Association for Computational Linguistics: NAACL 2024},
  pages={1519--1528},
  year={2024}
}

@article{liu2025differentially,
  title={Differentially private low-rank adaptation of large language model using federated learning},
  author={Liu, Xiao-Yang and Zhu, Rongyi and Zha, Daochen and Gao, Jiechao and Zhong, Shan and White, Matt and Qiu, Meikang},
  journal={ACM Transactions on Management Information Systems},
  volume={16},
  number={2},
  pages={1--24},
  year={2025},
  publisher={ACM New York, NY}
}

@inproceedings{mcmahan2017communication,
  title={Communication-efficient learning of deep networks from decentralized data},
  author={McMahan, Brendan and Moore, Eider and Ramage, Daniel and Hampson, Seth and y Arcas, Blaise Aguera},
  booktitle={Artificial intelligence and statistics},
  pages={1273--1282},
  year={2017},
  organization={Pmlr}
}

@article{hu2022lora,
  title={Lora: Low-rank adaptation of large language models.},
  author={Hu, Edward J and Shen, Yelong and Wallis, Phillip and Allen-Zhu, Zeyuan and Li, Yuanzhi and Wang, Shean and Wang, Liang and Chen, Weizhu and others},
  journal={Iclr},
  volume={1},
  number={2},
  pages={3},
  year={2022}
}

@article{guo2024selective,
  title={Selective aggregation for low-rank adaptation in federated learning},
  author={Guo, Pengxin and Zeng, Shuang and Wang, Yanran and Fan, Huijie and Wang, Feifei and Qu, Liangqiong},
  journal={arXiv preprint arXiv:2410.01463},
  year={2024}
}

@inproceedings{song2013stochastic,
  title={Stochastic gradient descent with differentially private updates.},
  author={Song, Shuang and Chaudhuri, Kamalika and Sarwate, Anand D and others},
  booktitle={GlobalSIP},
  pages={245--248},
  year={2013}
}

@inproceedings{bassily2014private,
  title={Private empirical risk minimization: Efficient algorithms and tight error bounds},
  author={Bassily, Raef and Smith, Adam and Thakurta, Abhradeep},
  booktitle={2014 IEEE 55th annual symposium on foundations of computer science},
  pages={464--473},
  year={2014},
  organization={IEEE}
}

@inproceedings{abadi2016deep,
  title={Deep learning with differential privacy},
  author={Abadi, Martin and Chu, Andy and Goodfellow, Ian and McMahan, H Brendan and Mironov, Ilya and Talwar, Kunal and Zhang, Li},
  booktitle={Proceedings of the 2016 ACM SIGSAC conference on computer and communications security},
  pages={308--318},
  year={2016}
}

@article{yin2021comprehensive,
  title={A comprehensive survey of privacy-preserving federated learning: A taxonomy, review, and future directions},
  author={Yin, Xuefei and Zhu, Yanming and Hu, Jiankun},
  journal={ACM Computing Surveys (CSUR)},
  volume={54},
  number={6},
  pages={1--36},
  year={2021},
  publisher={ACM New York, NY, USA}
}

@article{li2020federated,
  title={Federated learning: Challenges, methods, and future directions},
  author={Li, Tian and Sahu, Anit Kumar and Talwalkar, Ameet and Smith, Virginia},
  journal={IEEE signal processing magazine},
  volume={37},
  number={3},
  pages={50--60},
  year={2020},
  publisher={IEEE}
}

@article{zhang2026instruction,
  title={Instruction tuning for large language models: A survey},
  author={Zhang, Shengyu and Dong, Linfeng and Li, Xiaoya and Zhang, Sen and Sun, Xiaofei and Wang, Shuhe and Li, Jiwei and Hu, Runyi and Zhang, Tianwei and Wang, Guoyin and others},
  journal={ACM Computing Surveys},
  volume={58},
  number={7},
  pages={1--36},
  year={2026},
  publisher={ACM New York, NY}
}

@article{kang2024federated,
  title={Federated low-rank adaptation with differential privacy over wireless networks},
  author={Kang, Tianqu and Wang, Zixin and He, Hengtao and Zhang, Jun and Song, Shenghui and Letaief, Khaled B},
  journal={arXiv preprint arXiv:2411.07806},
  year={2024}
}

@article{zhang2023lora,
  title={Lora-fa: Memory-efficient low-rank adaptation for large language models fine-tuning},
  author={Zhang, Longteng and Zhang, Lin and Shi, Shaohuai and Chu, Xiaowen and Li, Bo},
  journal={arXiv preprint arXiv:2308.03303},
  year={2023}
}

@article{li2025uni,
  title={Uni-LoRA: One Vector is All You Need},
  author={Li, Kaiyang and Han, Shaobo and Su, Qing and Li, Wei and Cai, Zhipeng and Ji, Shihao},
  journal={arXiv preprint arXiv:2506.00799},
  year={2025}
}

\end{document}